\documentclass[a4paper,11pt]{article}
\pdfoutput=1 

\usepackage{jcappub} 

\usepackage[T1]{fontenc} 
\usepackage{mathtools}
\usepackage{amsthm}
\usepackage{bm}
\usepackage{graphicx,epsfig}
\usepackage{ulem}
\usepackage{xfrac}
\usepackage{color}
\usepackage{cancel}
\usepackage{bigints}

\usepackage[mathscr]{euscript}
\usepackage[T1]{fontenc}
\usepackage[latin9]{inputenc}
\usepackage{verbatim}
\usepackage{textcomp}
\usepackage{amsmath}
\usepackage{graphicx}
\usepackage{amssymb}
\usepackage{esint}

\usepackage{mathrsfs}

\usepackage{soul}
\usepackage{hyperref}
{
 \definecolor{BLACK}{gray}{0}
 \definecolor{WHITE}{gray}{1}
 \definecolor{RED}{rgb}{1,0,0}
 \definecolor{GREEN}{rgb}{0,1,0}
\definecolor{dgreen}{rgb}{.1,.6,.1}
\definecolor{BLUE}{rgb}{0,0,1}
 \definecolor{CYAN}{cmyk}{1,0,0,0}
 \definecolor{MAGENTA}{cmyk}{0,1,0,0}
 \definecolor{YELLOW}{cmyk}{0,0,1,0}
 \definecolor{aw}{rgb}{0.2,0.5,0.75}
  }
  \hypersetup{
    colorlinks,%
    citecolor=blue,%
    filecolor=blue,%
    linkcolor=magenta,%
    urlcolor=blue
}

\definecolor{MyDarkRed}{rgb}{0.71,0.14,0.07}

\usepackage{mathtools}

\newcommand{\mathsout}[1]
{\bgroup\mathchoice
  {\sbox0{$\displaystyle{#1}$}%
    \usebox0\hspace{-\wd0}%
    \rule[0.5\ht0-0.5\dp0-.5pt]{\wd0}{1pt}}%
  {\sbox0{$\textstyle{#1}$}%
    \usebox0\hspace{-\wd0}%
    \rule[0.5\ht0-0.5\dp0-.5pt]{\wd0}{1pt}}%
  {\sbox0{$\scriptstyle{#1}$}%
    \usebox0\hspace{-\wd0}%
    \rule[0.5\ht0-0.5\dp0-.5pt]{\wd0}{1pt}}%
  {\sbox0{$\scriptscriptstyle{#1}$}%
    \usebox0\hspace{-\wd0}%
    \rule[0.5\ht0-0.5\dp0-.5pt]{\wd0}{1pt}}%
\egroup}

\newcommand{\ba}{\begin{eqnarray}}
\newcommand{\ea}{\end{eqnarray}}

\newcommand{\bse}{\begin{subequations}}
\newcommand{\ese}{\end{subequations}}

\newcommand{\bbq}{\begin{quote}}
\newcommand{\eeq}{\end{quote}}

\newcommand{\CR}{{\cal{R}}}

\newcommand{\A}{{\cal{A}}}
\newcommand{\F}{{\cal{F}}}

\newcommand{\dd}{{\hbox{d}}}

\def\t#1#2#3{#1^{\if#2- \else #2 \fi}_{\if#2- \else \:\: \fi #3}}

\def\tdot#1#2#3{\dot{#1}^{\if#2- \else #2 \fi}_{\if#2- \else \:\: \fi #3}}
\def\tddot#1#2#3{\ddot{#1}^{\if#2- \else #2 \fi}_{\if#2- \else \:\: \fi #3}}
\def\ttilde#1#2#3{\tilde{#1}^{\if#2- \else #2 \fi}_{\if#2- \else \:\: \fi #3}}

\def\ba#1{\boldsymbol{\eta}^{#1}}

\font\bigastfont=cmr10 scaled \magstep 2
\def\bdot{\hbox{\bigastfont .}}
\newcommand{\dotaverage}[1]{\left\langle #1 \right\rangle^{\bdot}_\cD}

\newcommand{\cD}{{\cal D}}

\newcommand{\average}[1]{\left\langle #1 \right\rangle_\cD}

\newcommand{\CQ}{{\cal Q}}
\newcommand{\inI}{{\rm I}}
\newcommand{\inII}{{\rm II}}
\newcommand{\inIII}{{\rm III}}

\newcommand{\z}{z}
\newcommand{\y}{y}
\newcommand{\x}{x}
\newcommand{\rv}{\boldsymbol{r}}

\newcommand{\rmd}{{\rm d}}

\newcommand{\CA}{{\cal A}}

\newcommand{\CI}{{\cal I}}

\newcommand{\laverage}[1]{\left\langle #1 \right\rangle_{\cD_{\rm \bf i}}}

\newcommand{\aaverage}[1]{\left\langle #1 \right\rangle_{\CI}}

\newcommand{\initial}[1]{{#1_{\rm \bf i}}}

\title{On the maximum volume of collapsing structures}

\author[a,1]{Jan J. Ostrowski \note{Corresponding author.}}
\author[a,2]{Ismael Delgado Gaspar,\note{Corresponding author.}}

\affiliation[a]{Department of Fundamental Research, National Centre for Nuclear Research, Pasteura 7, 02-093 Warszawa, Poland}

\emailAdd{Jan.Jakub.Ostrowski@ncbj.gov.pl}
\emailAdd{Ismael.DelgadoGaspar@ncbj.gov.pl}

\abstract{
In many cosmological models, including the $\Lambda$CDM concordance model, there exist theoretical upper bounds on the size of collapsing structures. The most common formulations in the literature refer to a turnaround radius in spherical symmetry or a turnaround surface, defined as the zero-expansion boundary separating the outer Hubble flow from the inner flow of a collapsing fluid. In order to access a generic scenario, we propose an improvement of this cosmological test in terms of the maximum volume of the cosmological structures, which is equivalent to a zero-averaged expansion -- instead of the zero-local expansion. By combining the Lagrangian perturbations method and the scalar averaging of Einstein's equations, we obtain a maximum volume for a collapse model without any restricting symmetries. 
We compare this result with some exact, inhomogeneous solutions and discuss further potential developments.
}

\begin{document}
\maketitle
\flushbottom

\section{Introduction}
\label{sec:intro}

Cosmological large-scale structures can, in principle, provide robust tests of cosmological models. Due to the structures' morphological and dynamical complexity, such tests are required to cover a broad set of initial conditions and dynamical scenarios. In this context, the whole branch of standard tests considers the statistical properties of the density field, e.g.~\cite{Eisenstien2005},~\cite{Feldman1994}, \cite{Chuang2016},~\cite{Ballinger1996}. 

There exists, however, another complementary approach that does not aim at determining the parameters of any given cosmological model but sets certain bounds on observables and thus may be used to falsify or undermine a concrete theory (see, e.g.~\cite{Holz2012},~\cite{Clarkson2008}). 
One of such tests is based on the turnaround surface, i.e., the sphere of zero-expansion surrounding a collapsing structure marking the surface where the Hubble expansion and the gravitational attraction of the structure cancel each other out. However, this formulation is a bit unfortunate as it suggests that there are two competing physical processes: expansion and contraction, while according to general relativity, both are just manifestations of the evolution of the spatial metric (from the practical point of view, especially in the perturbation theory context, it is sometimes helpful to introduce this artificial division into the background and peculiar motion, keeping in mind that there is nothing fundamental behind this distinction). This method was proposed, e.g., in~\cite{Pavlidou2014}, where authors derive an upper limit on the radius of the zero-expansion sphere, i.e., its maximum value, at the epoch of decoupling from the Hubble flow, that can be regarded as a maximum value obtainable by any spherically symmetric structure in the $\Lambda$CDM model. 

In~\cite{Pavlidou2014}, three different methods are presented to derive the maximum turnaround radius 
\begin{equation}
\label{eq:Rmax}
    R_{max} = \left(\frac{3GM}{\Lambda}\right)^{1/3} \;,
\end{equation}
in the $\Lambda$CDM model (setting the speed of light to one).
These methods include using the geodesic equations in the Einstein-de Sitter space-time, the standard perturbation theory, and the non-linear evolution of an over-densities described by the Friedmann equations. All of the above methods treat a spherical, homogeneous over-densities and thus are very restricted. Some authors generalized this approach using the quasi-spherical models or the standard perturbation theory~\cite{Giusti2021},~\cite{DelPopolo2020}. A possible violation of the~\eqref{eq:Rmax} bound was reported in~\cite{Lee2015-2}. However, as pointed out in~\cite{Hansen2020}, determination of the zero-expansion surface can be observationally cumbersome and prone to errors; thus, another similar test based on a different principle has the potential to become very useful.

In this paper, we present a substantial improvement and
modification of the zero-expansion surface idea by extending the domain
of applicability to the non-spherical collapse. In the mean time, we define
a different condition for the turnaround. Namely, the maximum volume
that the structure can attain depending on the cosmological mode
 With this improvement, we are able to derive an upper bound for a generic collapse scenario. The main motivations for the use of a maximum volume
method follows from two facts: (i) it considerably simplifies the averaged
Hamiltonian constraint due to the corresponding zero-averaged expansion,
(ii) it is, in principle, detectable. We present this feature with examples, including the relativistic Zel'dovich approximation (RZA) and some exact relativistic solutions.
  
This article is organized as follows. First, in Sec.~\ref{sec:BackAveEqns}, we introduce the mathematical tools and concepts necessary to develop our approach, i.e., the scalar averaging of Einstein's equations and the relativistic Zel'dovich approximation. The following section, Sec.~\ref{sec:result}, is dedicated to the derivation of our main result, the maximum (zero-averaged expansion) volume of the cosmic structure as a function of redshift, mass, and the background cosmology, supplemented with some numerical solutions. Our results are discussed in Sec.~\ref{sec:discussion-compa}, where we also look at some illustrative exact cases. Section~\ref{sec:conclusions} concludes the paper with some final remarks and an outlook for future research.


\section{Spatially averaged Einstein's equations and the relativistic Zel'dovich approximation}\label{sec:BackAveEqns}
 
In the context of extended objects in general relativity, such as cosmological structures, it is advantageous to utilize the scalar averaging scheme (see e.g., \cite{Buchert2000-2}). This formalism takes the scalar parts of Einstein's equations in $3+1$ form and applies to them a spatial averaging operator. Since the spatial averaging and time evolution do not commute, the resulting equations exhibit an additional term, the backreaction. There has been a considerable debate on the nature of backreaction in general relativity, see e.g. \cite{Buchert2015} and \cite{Green2011}. In the context of this paper, however, the use of averaged equations seems to be non-controversial as it is a part of the standard cosmological paradigm that the global behaviour of the Universe is treated relativistically, while the structures obey the laws of Newtonian physics.   
More specifically, in both relativistic and Newtonian cases, the governing averaged equations are similar apart from being subjected to different interpretations, e.g., in the general relativistic context we use the extrinsic curvature tensor, while in the Newtonian case we speak about the velocity gradient. In the following we will use the relativistic equations; however, the main result of this paper remains unaffected by this choice. 

We continue this section by recalling the basic structure of scalar averaging.
Within the fluid-orthogonal foliation of space-time, we can define the spatial averages of the scalar parts of Einstein's equations with irrotational dust chosen as the source. We define the averaging operator over a compact spatial domain $\cD$ as:
\begin{equation}
  \average{\CA}=\frac{1}{V_{\cD}}\int_{\cD} \,\rmd \mu\
  \CA\ \;, \, V_{\cD} = \int_{\cD} \,\rmd \mu\ \;,
\end{equation}
where $\rmd \mu$ is the Riemannian volume element. 

The Friedmann equations of the Friedmann-Lema\^\i tre-Robertson-Walker (FLRW) model find their generalization in an inhomogeneous model through the volume expansion and acceleration equations ~\cite{Buchert2000-2}. We define the domain-dependent scale factor as 
\begin{equation}
a_{\cD}=\left(V_{\cD}/V_{\cD_i}\right)^{\frac{1}{3}} \ ; \quad V_{\cD_i}=V_{\cD}(t_i) \ .
\end{equation}
The spatially averaged energy constraint (the Hamiltonian constraint), the Raychaudhuri's equation, and mass conservation form a set of exact equations which include the extrinsic curvature invariants collected in the  backreaction term:

\bse\label{AppEq:SetAvedEqsQd}
\begin{eqnarray}
3 \left(\frac{\dot{a}_\cD}{a_\cD}\right)^2=\Lambda+ 8 \pi \average{\varrho} - \frac{1}{2}\average{{}^{(3)}\CR}-\frac{1}{2}\CQ_{\cD} \ ; \qquad
\label{AppEq:BackR-da}
\\
3 \frac{\ddot{a}_\cD}{a_\cD}= \Lambda -4 \pi \average{\varrho} + \CQ_{\cD} \ ;  \qquad\label{AppEq:BackR-dda}
\\
\dotaverage{\varrho}= -3 \frac{\dot{a}_\cD}{a_\cD} \average{\varrho} \ , \qquad\label{AppEq:BackR-rho}
\end{eqnarray}
\ese
where the kinematical backreaction term reads:
\begin{equation}\label{Eq:QdInv-a}
\CQ_{\cD}\equiv 2 \average{\inII}-\frac{2}{3} \average{\inI}^2  \ .
\end{equation}
$\average{\inI}$ and $\average{\inII}$ are the averaged principal invariants of (minus) the extrinsic curvature tensor, the expansion tensor $\Theta_{ij}$:
\begin{equation}
\label{eq:invariants}
    \Theta_{ij} = \frac{1}{2}\dot{g}_{ij} \;,\; \inI = \mathrm{tr}\left(\Theta_{ij}\right)
\;,\;\inII = \frac{1}{2}\left(\mathrm{tr}\left(\Theta_{ij}\right)^2-\mathrm{tr}\left(\left (\Theta_{ij}\right)^2\right)\right)\;,
\end{equation}
and $\average{{}^{(3)}\CR}$ is the averaged scalar curvature. This system of equations is undetermined. In the series of papers \cite{Buchert2012,Buchert2013,Alles2015,Buchert2017}, the authors propose a closure condition derived from the relativistic version of the Zel'dovich approximation. The main idea consists of expressing the $3+1$ Einstein's equations using only one tensorial variable, the co-frame, and postulating its particular form as a perturbation with respect to the background. No further linearization is performed, i.e., the perturbed co-frame is used to evaluate all the relevant fields without truncating the higher order terms in functionals of the co-frames. This line of reasoning leads to the closure condition of the averaged equations with the backreaction term reading: 
\begin{equation}\label{Eq:QdInv-b}
\CQ_{\cD}\equiv 2 \average{\inII}-\frac{2}{3} \average{\inI}^2  
=\frac{\aaverage{\ddot{{\mathfrak{J}}}}}{\aaverage{{\mathfrak{J}}}}-\frac{\ddot{\xi}}{\dot{\xi}}
\frac{\aaverage{\dot{{\mathfrak J}}}}{\aaverage{{\mathfrak J}}}-\frac{2}{3}\left(\frac{\aaverage{\dot{{\mathfrak J}}}}{\aaverage{{\mathfrak J}}}\right)^{2}
\ .
\end{equation}
In the above equation,  ${\mathfrak J}$ is the peculiar-volume deformation: 
\begin{equation}
{\mathfrak J} = 1+\xi(t)\inI_i +\xi^2(t)\inII_i +\xi^3(t)\inIII_i \;,
\end{equation}
with the initial values of invariants \eqref{eq:invariants} (subscript $i$), the third invariant reading: $\inIII = \mathrm{det}\left(\Theta_{ij}\right)$,
and $\xi\left(t\right)$ being the background-dependent growth function defined by
\begin{equation}
  \xi(t):=\lbrack q(t)-q(\initial t)\rbrack/{\dot{q}(\initial t)}\ ,
\label{eq:def-xi}
\end{equation}
where $q(t)$ obeys
\begin{equation}\label{eq:evolution-q}
  \ddot{q}(t)+2H\dot{q}(t)-4\pi \varrho_H q(t)=0\ ,
\end{equation}
i.e., the same equation that leads to the growing and decaying modes in the linear cosmological perturbation theory.
In \ref{Eq:QdInv-b} we used the formal average operator:
\begin{equation}
  \aaverage{\CA}=\frac{1}{V_{\initial\cD}}\int_{\cD} \,\rmd^{3}X\
  \CA\ ,
\end{equation}
where $X$ stands for the Lagrangian coordinates (see \cite{Buchert2013} for details).
With this definition, the volume Hubble expansion rate can be expressed as follows
\begin{equation}\label{Eq:HDDef}
H_\cD=\frac{\dot{a}_\cD}{a_\cD} = H+\frac{1}{3}\frac{\aaverage{\dot{\mathfrak J}}}{\aaverage{\mathfrak J}}\ .
\end{equation}
A similar relationship can be found for $\ddot{a}_\cD/a_\cD$:
\begin{equation}\label{Eq:ddacDDef}
\frac{\ddot{a}_\cD}{a_\cD} + 2 H_\cD^2 = 
\frac{\ddot{a}}{a} + 2 H^2
+2 H\frac{\aaverage{\dot{\mathfrak J}}}{\aaverage{\mathfrak J}}
+\frac{1}{3}\frac{\aaverage{\ddot{\mathfrak J}}}{\aaverage{\mathfrak J}} \ .
\end{equation}
In the following section we will use the tools briefly sketched here to develop a general expression for the turnaround volume.  
%


\section{Turnaround condition}
\label{sec:result}

In the case of realistic structures, it is not always easy to define the turnaround epoch since, in inhomogeneous spacetimes, each fluid element follows different type of evolution and decouples from the Hubble flow at different times. This problem is absent in the simplified homogeneous case where the turnaround is synchronous. Throughout this paper we will use the definition proposed in \cite{Roukema2019} based on the vanishing of the average Hubble expansion:
\begin{equation}\label{Eq:TADef}
\mbox{Turnaround:} \qquad 3H_\cD=\average{\Theta}=\frac{\dot{V}_{\cD}}{V_{\cD}}=0 \implies \dot{V}_{\cD}=0 \ .    
\end{equation}
This criterion constitutes the most direct generalization of the zero expansion surface. The standard representation of a homogenous dust ball reaching a maximum radius (volume) and then recollapsing is recast in terms of the averaged fields. 
In our case, the turnaround is marked by the time when the volume ($\dot{V}_\cD=0$) or the characteristic scale of the configuration ($\ell \sim V^{1/3}$) reach their maximum.

Our definition of the turnaround is general and can be used in scenarios where RZA is no longer applicable, e.g., in the case of exact cosmological solutions that do not admit a clear distinction between the background and peculiar fields. In fact, this kind of study offers an important insights into the large-scale structure formation in the non-linear regime. In Sec.~\ref{sec:beyond}, we will discuss two illustrative toy models based on Szekeres solutions; in this section we will solely focus on the RZA approach.

The turnaround condition, Eq.~\eqref{Eq:TADef}, and equations~\eqref{Eq:HDDef},~\eqref{Eq:ddacDDef}, and~\eqref{Eq:QdInv-a} lead to
\bse\label{SubEqs:HDddAdQD}
\begin{eqnarray}
\frac{\aaverage{\dot{\mathfrak J}}}{\aaverage{\mathfrak J}}=-3H \ ,
\\
\frac{\ddot{a}_\cD}{a_\cD}  =  
-\frac{9}{2} H^2
+\frac{\Lambda}{2} 
-\frac{k}{2 a^2}
+\frac{1}{3}\frac{ \aaverage{\ddot{\mathfrak J}}}{\aaverage{{\mathfrak J}}}\ , 
\\
\CQ_{\cD}
=
\frac{\aaverage{\ddot{{\mathfrak J}}}}{\aaverage{{\mathfrak J}}}
+3H\frac{\ddot{\xi}}{\dot{\xi}}
-6 H^{2} \ .
\end{eqnarray}
\ese
To relate the average and the background densities (at the turnaround), we first insert~\eqref{SubEqs:HDddAdQD} into~\eqref{AppEq:BackR-dda}; then, considering only  the growing solution to $q$, we obtain  
\begin{equation}\label{Eq:DenFinal}
\frac{ \average{\varrho}}{\varrho_H}=  1 + 3 \left[H\bigg/\left(\frac{\dot{q}}{q}\right)\right] \ , 
\quad \mbox{with} \quad
q\simeq q_+\propto \frac{\dot a}{a} \int_0^a \dot{\hat{a}}^{-3} \dd \hat{a} \ ,
\end{equation}
which for an EdS model (i.e., $q=a$) reduces to $\average{\varrho}/\varrho_H= 4$, confirming the result obtained by a slightly different method in \cite{Ostrowski2019} and \cite{Roukema2019}. Note that $H$, $q$ and $\dot q$ are understood to be evaluated at the turnaround. 
Equation~\eqref{Eq:DenFinal} involves the structure mass and volume in addition to the background functions, which establishes the maximum volume without specifying the initial conditions and regardless of the details of the collapse. 

As a final remark in this section, we would like to point out that this derivation differs from the one in \cite{Roukema2019} and \cite{Ostrowski2019}. This is due to the subtle ambiguity inherent to the RZA, where two definitions of the domain-dependent scale factor, the kinematic and dynamical, are in principle possible. In our derivation, we use the kinematic definition, which is usually considered slightly inferior; however, since we are aiming at describing early stages of the mildly non-linear regime (up to a turnaround), both definitions are equally acceptable (see \cite{Buchert2000} for details and the direct comparison of these definitions).


\subsection{Turnaround ``radius''}

To pave the relation with other treatments in the literature, we supplement this section with an estimation of the characteristic scale of the cosmic structure at the turnaround.  
At the time of the last scattering, the density distribution  corresponds to fluctuations ($\Delta\rho_H(\initial t)$) around the density value predicted by the Friedmann model ($\rho_H(\initial t)$). These fluctuations are usually several orders of magnitude smaller than $\rho_H(\initial t)$; hence, 
\begin{equation}\label{Eq:AverRhotApp}
\average{\varrho}=\frac{\laverage{\varrho(\initial t)}}{a_{\cD}^{3}}=\frac{\int_{\cD_i} \,\rmd^{3}X \left(\rho_H(\initial t)+\Delta\rho_H(\initial t)\right)}{a_{\cD}^{3}V_{\initial{\cD}}}\simeq \frac{\rho_H(\initial t)}{a_{\cD}^{3}} \ .
\end{equation}
Substituting~\eqref{Eq:AverRhotApp} into~\eqref{Eq:DenFinal}, we find a relationship between the background and average scale factors: 
\begin{equation}
    \left(\frac{a}{a_{\cD}}\right)^{3} \simeq  1 + 3 \left[H\bigg/\left(\frac{\dot{q}}{q}\right)\right] \ .
\end{equation}
This expression can be rewritten as an estimate of the volume at the turnaround
\begin{equation}\label{Eq:VolEstimt}
V_{\cD}(t_{ta}) = \frac{M}{\rho_H\left(1+3H\left(\frac{\dot{q}}{q}\right)^{-1}\right)} \ ,
\end{equation}
where $M=M(t_i)\simeq \varrho_H(t_i) V_{\cD_i}$ is the (conserved) mass of the structure. 

The structure volume $V_{\cD}(t_{ta})$ could be associated with its characteristic length or radius: $\ell^3(t_{ta})\sim V(t_{ta})$. 

However, in an anisotropic (pancake or filamentary) collapse, the length scale in one eigen-direction differs considerably from the others and thus we note that the characteristic scale has only a limited applicability i.e. the (quasi-)spherical models.


\subsection{Some numerical results}

To examine the predictions of formulas~\eqref{Eq:DenFinal} and~\eqref{Eq:VolEstimt}, it is convenient to express them in terms of the redshift and the present-day cosmic parameters (\cite{Planck2020}): 
$\Omega_{m0}= 0.315$, 
$\Omega_{\Lambda0}=1-\Omega_{m0}$,
$H_0 = 67.4 \,\mbox{km} \,\mbox{s}^{-1} \mbox{Mpc}^{-1}$; also, we will formally allow a non-vanishing spatial curvature, $\Omega_{k0}$.
Then, Eq.~\eqref{Eq:DenFinal} takes the following form:
\begin{equation}\label{Eq:RatioRhosZ}
\frac{ \average{\varrho}}{\varrho_H}= 1+3 \zeta(z)^{-1} \ ,
\end{equation}
with
\begin{eqnarray}
 \zeta(z)&:=&\frac{\dot{q}}{q H}
 = -\frac{3}{2}\frac{\Omega_{m0}(1+z)^3}{E^2(z)}-\Omega_{k0} \frac{(1+z)^2 }{E^2(z)} +\frac{(1+z)^2}{E^3(z)}
 \left(\int_z^\infty \dd z^{\prime}\frac{(1+z^{\prime})}{E^3(z^{\prime})}\right)^{-1} , 
 \label{Eq:Zetaofz}
 \qquad
 \\
 E^2(z)&:=&\frac{H^2}{H_0^2}=\left(\Omega_{m0}\left(1+z\right)^3+\Omega_{k0}(1+z)^2+\Omega_{\Lambda0}\right) \ .
\end{eqnarray}

 Similarly, Eq.~\eqref{Eq:VolEstimt} leads to an approximate expression for the dimension of the structures reaching the turnaround at a particular $z$:
\begin{equation}\label{Eq:VolEstimZ}
V(z) \simeq \frac{1}{(1+z)^3}\left( \frac{M}{\varrho_{c0} \Omega_{m0} \left(1 + 3  \zeta(z)^{-1} \right)} \right)\ .
\end{equation}
As before, $M$ is the mass of the structure and $\varrho_{c0}=\frac{3 H_0^2}{8 \pi G}$ is the critical density. 

\begin{figure}[tbp]
\centering 
\includegraphics[width=.48\textwidth]{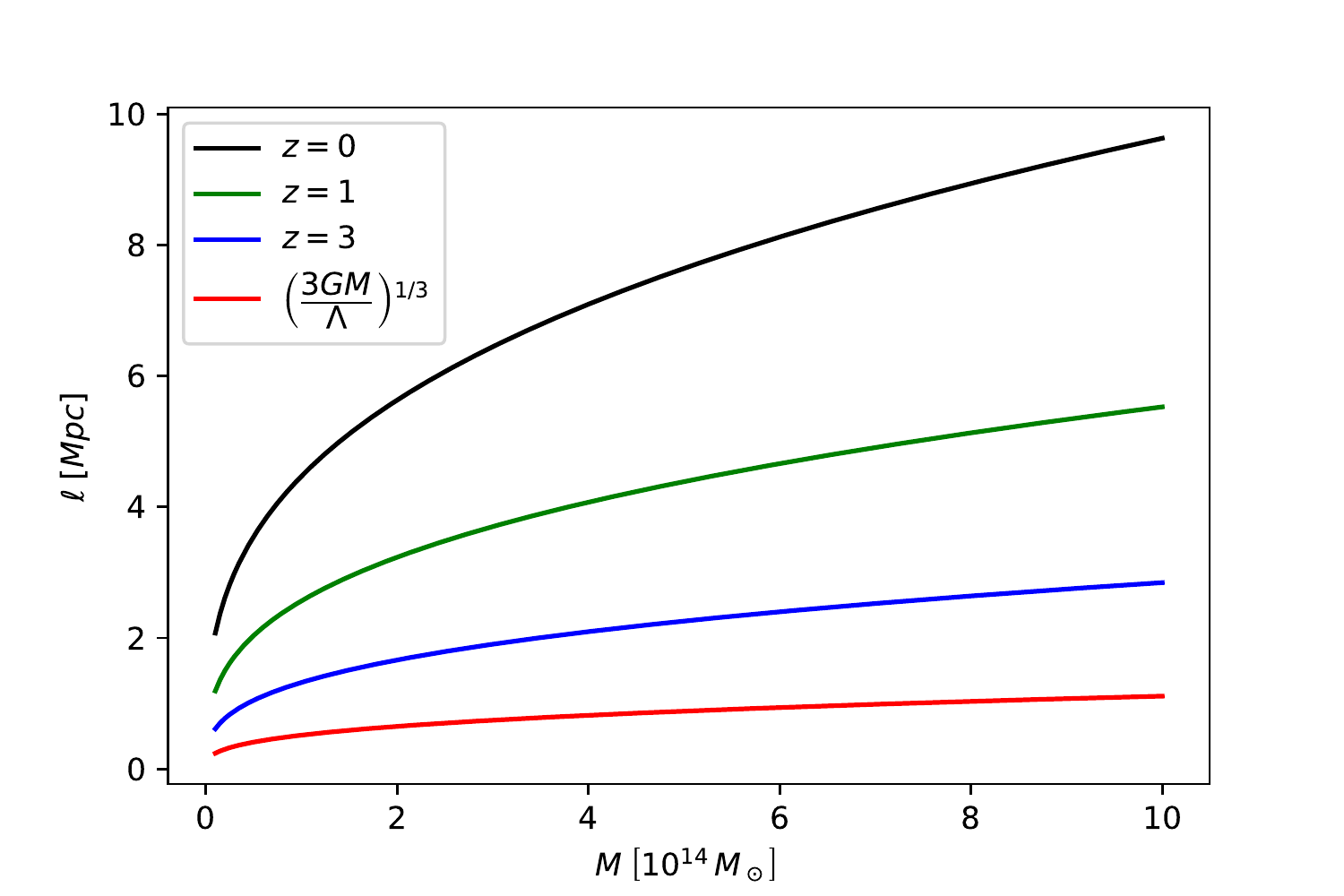}
\hfill
\includegraphics[width=.48\textwidth]{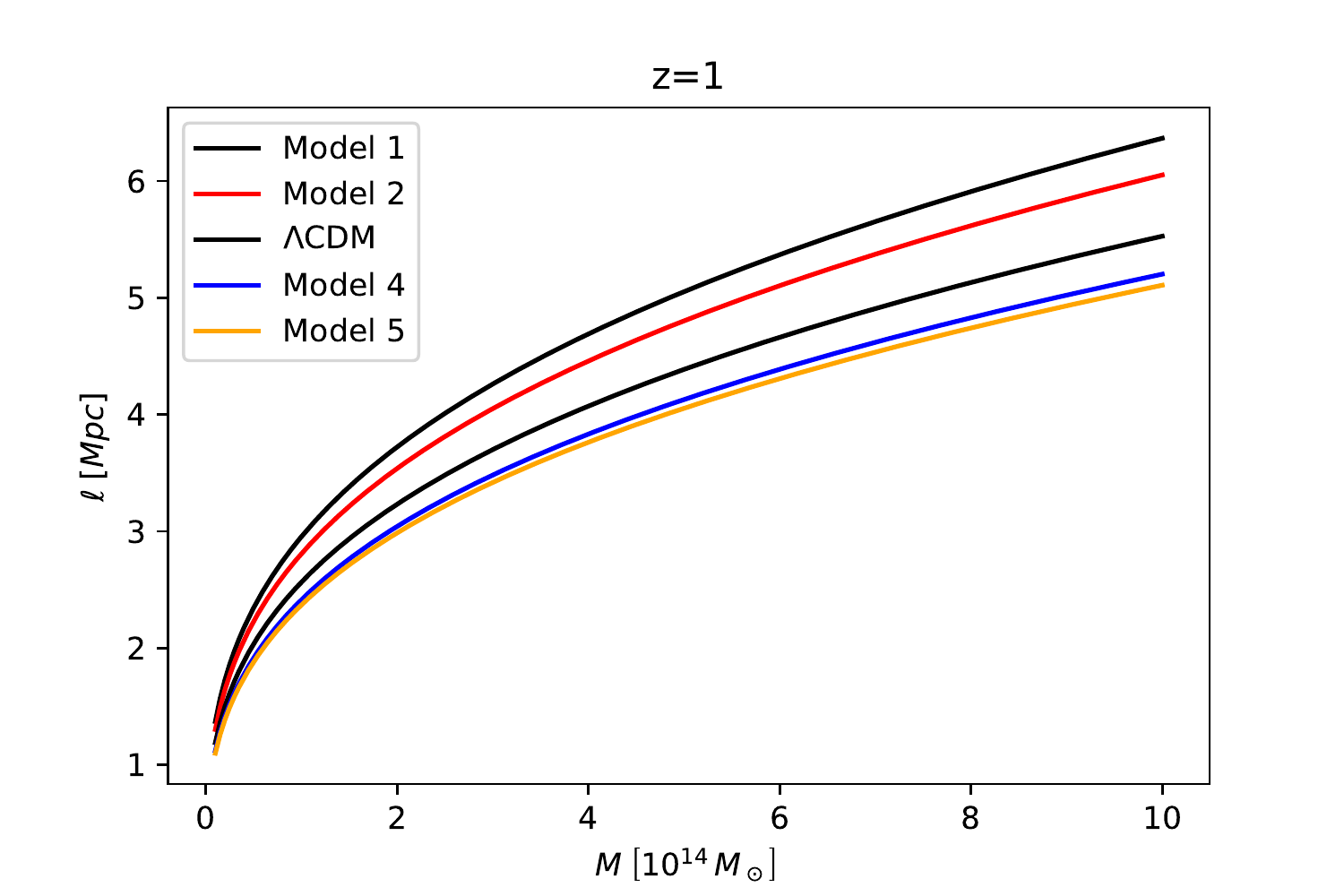}
\caption{\label{fig:RvsM} 
{\bf Characteristic  length as a function of the mass} 
for $z=0,1,3$ (left panel) and different cosmologies at $z=1$ (right panel).
In the latter the five models correspond to different values of $\Omega_{k0}$ at $z=0$: $\Omega_{k0}=\left[-1+\Omega_{m0},-0.5,0,0.5,1-\Omega_{m0}\right]$,
while
$\Omega_{m0}=0.315$ 
and 
$\Omega_{\Lambda}=1-\Omega_{m0}-\Omega^{(i)}_{k0}$.
}
\end{figure}

In Fig. \ref{fig:RvsM} (left panel), we observe the maximum length scales as a function of the mass in the $\Lambda$CDM cosmology. The zero expansion radius is plotted for comparison. As expected, the zero expansion surface is contained within the maximum volume region. As a proof-of-concept, we  plot the maximum radius at $z=1$ for different sets of cosmological parameters in the right panel. 
It is worth noticing that the discrepancies between different background models become substantial with the increase of the mass. 

\begin{figure}[tbp]
\centering 
\includegraphics[width=.48\textwidth]{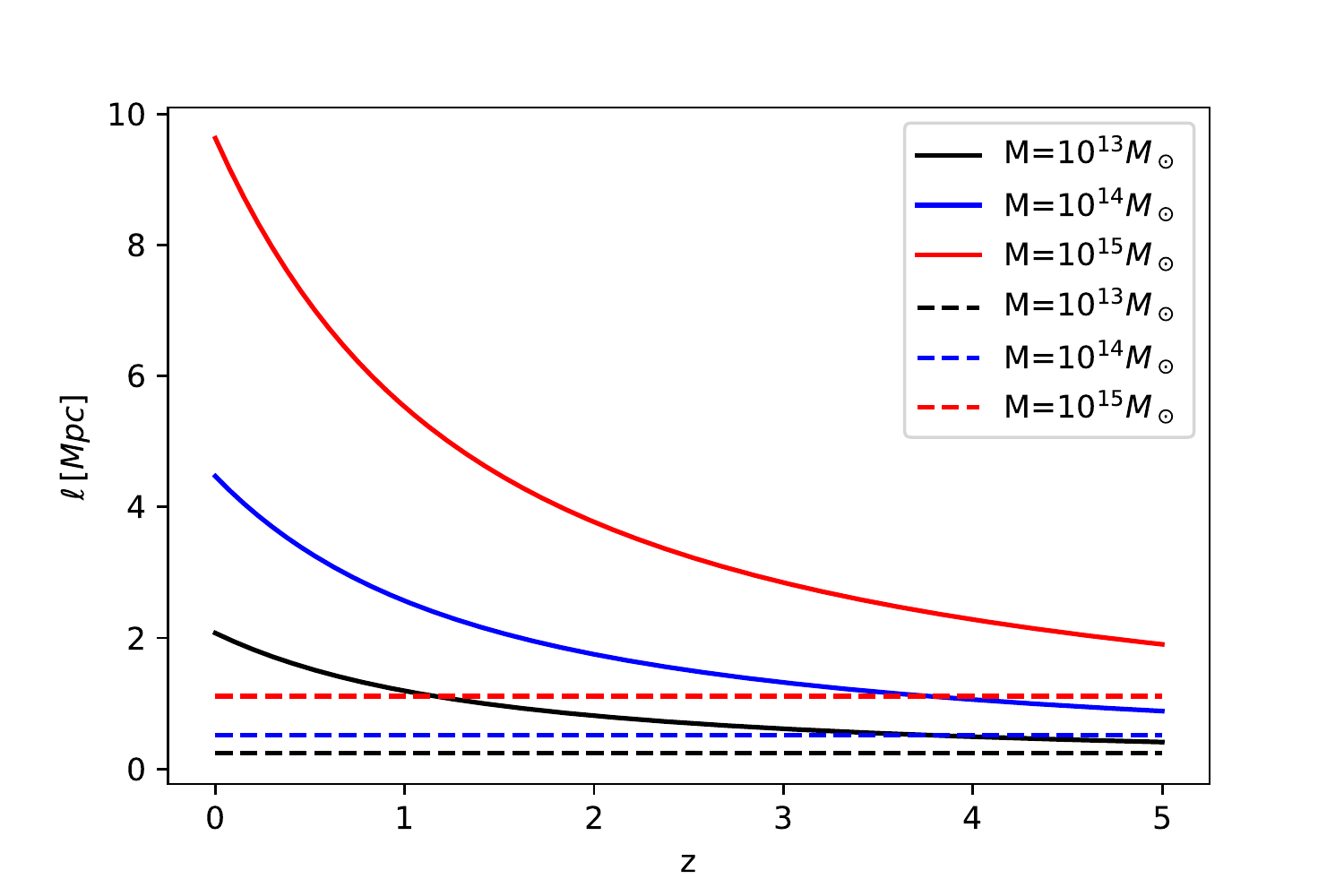}
\hfill
\includegraphics[width=.48\textwidth]{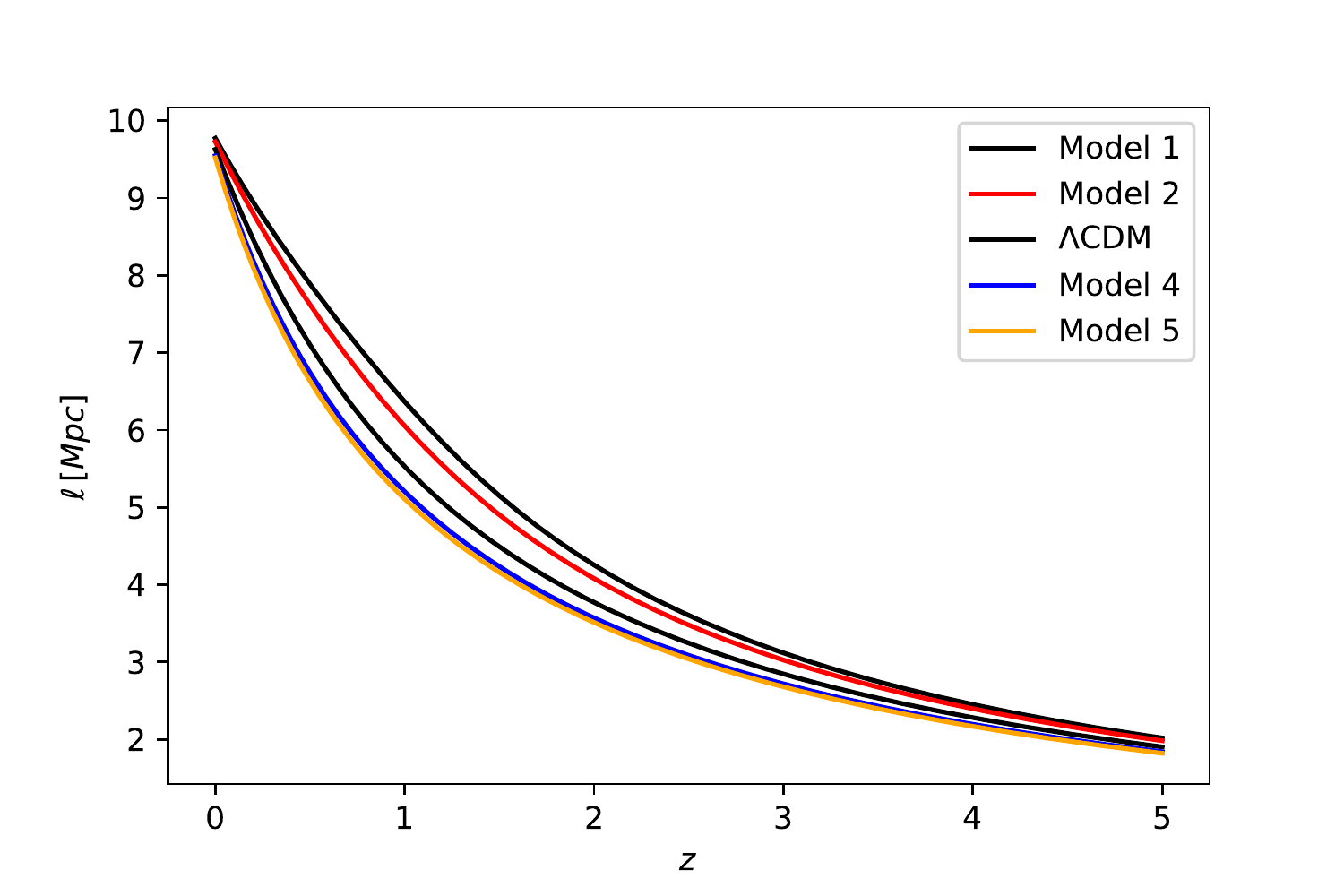}
\caption{\label{fig:RvsZ} 
{\bf Characteristic length at the turnaround as a function of the redshift}
for structures with $M=10^{13},10^{14},10^{15} M_\odot$ (left panel), and different cosmological models with $M=10^{15} M_\odot$ (right panel).
In the left panel, the lines $R=3MG/\Lambda$ for each value of the mass are displayed as a reference, and in the right panel,
the models are the same as in Fig.~\ref{fig:RvsM}.
}
\end{figure}

In the left panel of Fig. \ref{fig:RvsZ}, the redshift dependence of the maximum radius is plotted against the constant in time zero-expansion radius. 
These plots suggest that the maximum turnaround radius (maximum volume) can complement the zero-expansion radius test at lower redshifts as it becomes visibly less correlated. The right panel of the same figure  
confirms that the low redshift observations are the most sensible for testing different background models.

\subsection{Applications}
The turnaround epoch, as described in the previous sections, is a distinctive stage in the evolution of cosmological structures that can be used to test certain theories. In this section, we will briefly discuss two possible tests: observational and numerical. Formula \ref{Eq:VolEstimt} gives us the upper bound on the volume of any structure, during its collapse history, as a function of the mass, the background-dependent growth function and the scale factor. In principle, determination of these quantities is possible observationally, and a very similar approach was attempted in order to determine the zero-expansion surface (\cite{Lee2015}). The main difference is that to claim that any particular zero averaged expansion volume violates the upper bound we need to determine whether the still expanding region of the given structure will be a part of the collapsed object in the future. One possible method to obtain it is based on the velocity analysis of the outer, still expanding regions of the galaxy clusters and finding its correlation with the central structure \cite{Falco2014}. Alternatively, by estimating the mass of the galaxy cluster or super-cluster with the X-ray emission, velocity dispersion, gravitational lensing or Sunayev-Zel'dovich effect, we can use the caustics method (see e.g. \cite{Serra2011}) to determine whether the given tracer (galaxy) has a velocity below the escape velocity and thus is a part of the bound structure that has not yet virialized.

Another application of the maximum volume is related to the N-body simulations and the backreaction conjecture. Estimates of the backreaction have been so far addressed either within the fluid formalism:   e.g. \cite{Kazimierczak2018} and \cite{Macpherson2019} or within the N-body codes but on the global scale: \cite{Adamek2019}. With the maximum volume approach, we can test the predictions of the averaged scalar equations together with the RZA serving as a closure condition by looking at the scales of clusters and super-clusters in the N-body simulations. The construction of the RZA is well suited for such comparisons as it relies on the notion of background, a generic feature of the majority of the N-body architectures. In \cite{Kordikis2020}, a related examination was performed to find the zero-expansion radius.


\section{Discussion and comparison with exact cases}
\label{sec:discussion-compa}

In Sec.~\ref{sec:result}, we introduced the turnaround stage as marked by the vanishing of the average of the expansion scalar, furnishing the most direct generalization of the analogous definition in the standard spherical collapse model. The predictions coming from applying this definition reduce to well-known results in the homogeneous limit.

As a part of our approach, we are averaging out the solution's intrinsic anisotropy and inhomogeneity, and reducing all the relevant physical information to a single kinematical quantity, the volume. Such a description matches our intuition in a (quasi-)isotropic collapse but should not be  applied to a highly anisotropic pancake collapse where the structure is found collapsing along two directions but expanding along the third (see an explicit solution further below).

Some comments are in order about the applicability of the expressions for the density ratio and the estimation of the volume at the turnaround time. While our definition of the turnaround is general,  Eq.~\eqref{Eq:RatioRhosZ} and~\eqref{Eq:VolEstimZ} were obtained under the assumption that there exists a referential cosmological background and that the structures form from a growing mode of the solution of~\eqref{eq:evolution-q}. This relates the domain of applicability of our model to the one of the RZA, which is known to provide an accurate description of the structure formation in the mildly non-linear regime. Consequently, our model and its predictions are expected to be accurate for the turnaround of the galaxy clusters and super-clusters.


\subsection{Some exact models}
\label{sec:exactmodels}

In order to gain a better insight into the properties of the turnaround condition, we examine some exact illustrative solutions. First, we consider the Szekeres models of class II and interpret the solution as a fluctuation on the FLRW background. This constitutes the exact limit of the model presented in Sec.~\ref{sec:result}. Then, we relax the hypothesis of the coupling of the growing fluctuations to the background, and search for possible extensions of the current formalism.

\subsubsection{Szekeres class II as exact fluctuations on an FLRW background}
\label{sec:ExactSoltolRZA}

The RZA contains the Szekeres class II models as an exact subcase~\cite{RZA6}. In this sense, it is not surprising that the approximate expressions for the density ratio and volume at the turnaround become exact for the Szekeres universes. However, demonstration of this fact constitutes a necessary consistency test for our model. 

The line-element of Szekeres-II models can be cast as follows (see~\cite{GW} for the introduction of this parametrization and~\cite{kras1,kras2,InhomogModels} for a detailed analysis of the Szekeres solutions)
\begin{equation}\label{Eq:SzeMetricGW}
\dd s^2=-\dd t^2 + a^2  \left((\A(\rv)-\F(t,\z))^2 \dd \z^{2} + e^{2 \nu(\rv)}  \left( \dd \x^{2} + \dd \y^{2}\right) \right) \ ,
\end{equation}
where $\A$ and $e^\nu$ are functions of the comoving coordinates\footnote{
The metric functions $e^\nu$ and $\A$  are given by $e^\nu=\left(1+k_0(\x^2+\y^2)/4\right)^{-1}$, $\A=e^\nu \Big[ c_0(\z) \left(1-\frac{k_0}{4}(\x^2+\y^2)\right) + c_1(\z)\x +c_2(\z)\y \Big]-k_0\beta_{+}$ for $k_0=\pm 1$, and $\A=c_0(\z)+c_1(\z)\x+c_2(\z)\y-\beta_{+} (\z) (\x^2+\y^2)/2$ for $k_0=0$.
}. The conformal scale factor, $a(t)$, obeys a Friedmann-like equation:
\begin{equation}\label{Eq:FriedmannLikeEqn}
\dot{a}^2=-k_0 + \frac{2\mu}{a} +\frac{\Lambda}{3} a^2 \ ,
\end{equation}
with $\mu=\mbox{const.}$, $k_0=0,\pm1$ and
$\F(t,\z)=\beta_{+}(\z) q_{+}(t)+\beta_{-}(\z) q_{-}(t)$.
The functions $\beta_\pm$ are arbitrary, and $q_\pm (t)$ obey~\eqref{eq:evolution-q} but with $4\pi \varrho_H \rightarrow 3 \mu$.
As in Sec.~\ref{sec:result}, we will neglect the contribution of the decaying solution ($\beta_-=0$).

So far, this is only an ansatz; however, if the constant $\mu$ is identified as the initial value of the background density, $3 \mu=4\pi \varrho_H(t_i)$, the solution acquires an interpretation as an exact fluctuation evolving on the Friedmann background.

To obtain the relationship between the physical quantities at the turnaround, we first average the density and expansion scalar of the dust source,
 \begin{eqnarray}
\varrho(t,\rv) 
= \frac{\varrho_{H}(t_i)}{a^3}\left(\frac{\A}{\A-\beta\, q}\right)  \ , 
\quad
\mbox{and}
\quad \mathcal{H}=\frac{\dot{a}}{a} -\frac{1}{3}\frac{\beta_+ \dot{q}_+}{\A-\beta_+ q_+} \ ,
\end{eqnarray}
which gives:
\begin{eqnarray}
\average{\varrho}
=\varrho_{H}
\frac{1}{1-q_+ (\chi_{\mathcal{D}}/\upsilon_{\mathcal{D}})} \ ,
\quad
\mbox{and}
\quad
H_{\cD}=H-\frac{1}{3}\frac{\dot{q}_+ \chi_{\mathcal{D}}}
		                  {\upsilon_{\mathcal{D}}-q_+ \chi_{\mathcal{D}}} \ ,
\label{Eq:AveRhoSze}
\end{eqnarray}
with $ \chi_{\mathcal{D}}:=\int_\cD \beta_+ e^{2\nu} d^3 \rv$ and $\upsilon_{\mathcal{D}}:=\int_\cD\A e^{2\nu} d^3 \rv$.
Then, imposing the turnaround condition, $H_{\cD}=0$, solving it for $\chi_{\mathcal{D}}/\upsilon_{\mathcal{D}}$, and substituting the result into the first equation in~\eqref{Eq:AveRhoSze}, we obtain the same expression as in~\eqref{Eq:DenFinal}. However, unlike the equations obtained in Sec.~\ref{sec:result}, this relation is exact. Finally, Eq.~\eqref{Eq:VolEstimt} can be found by following the same steps as in the approximate case.


\subsubsection{Beyond the RZA}
\label{sec:beyond}

The existence of a cosmological background coupled to the inhomogeneities  has been one of our fundamental assumptions so far. However, there are physical situations where one can treat the inhomogeneities as separate universes or dispense with the existence of the background. The simplest example is the standard spherical collapse model, for which our definition of the turnaround leads to the same well-known results since $H_{\cD}=H$.

The collapse of a spherical and homogeneous overdensity, evolving separately from the background, can be easily generalized using the Szekeres models. Taking $k_0 = 1$ and $\Lambda=0$ in~\eqref{Eq:FriedmannLikeEqn}, the Einstein's equations have analytical solutions (as before, we are neglecting the decaying solution for $q$)~\cite{GW}: 
\begin{eqnarray}
t(\theta)=\mu\left(\theta-\sin\theta\right) \ ,
\quad
a(\theta)=\mu\left(1-\cos\theta\right) \ ,
\quad
q(\theta)=\frac{6 \mu}{a}\left(1-\frac{\theta}{2}\cot\frac{\theta}{2}\right)-1 \ .
\end{eqnarray}
In this model, Eq.~\eqref{Eq:FriedmannLikeEqn} does not describe the evolution of a physical background but should be interpreted as a convenient ansatz. 
 ---To have a non-FLRW model and avoid shell-crossing singularities the free functions should satisfy: $\beta<0$ and $\mu>0$.
Our definition of the turnaround leads to the second equation in~\eqref{Eq:AveRhoSze}, which implies that: 
\begin{equation}
\frac{\dot a}{a}=-\frac{1}{3}\frac{\dot{q}\, 
\int_\cD |\beta| e^{2\nu} d^3 \rv}{\int_\cD\A e^{2\nu} d^3 \rv-q \, \int_\cD |\beta| e^{2\nu} d^3 \rv}<0\;.
\end{equation}
Thus the turnaround is reached at a certain time while the structure is recollapsing along the directions $x$ and $y$, but expanding along $z$. 
This mechanism allows for significant differences between the characteristic scales of the structure along different eigendirections. Consequently, a description based on a unique characteristic length ($\ell\sim V^{1/3}$) might not be accurate.


\section{Conclusions}
\label{sec:conclusions}
The zero average expansion condition has been proposed to mark the turnaround epoch of a generic collapsing structure to which the dust description applies. The scalar averaged Einstein's equations together with the RZA closure condition were employed to model the evolution of spatially extended objects. In particular, the averaged Hamiltonian constraint with the average zero-expansion condition allows one to determine the maximum volume attainable by cosmological structures as a function of redshift, mass and background model, removing the dependence on the initial conditions. This formulation leads to two possible cosmological tests: the upper bound on the volume of collapsing objects which can be used to exclude particular background models, including the concordance $\Lambda$CDM model, and the N-body tests directly related to the backreaction conjecture on the scales of clusters and super-clusters (similar considerations within alternative theories of gravity are in principle possible). The observational determination of the volume of the structure can be supplemented with the additional information on the mass of the structure, estimated from the mass function of the galaxy clusters and super-clusters. This was done in \cite{Ostrowski2017} within the RZA context, only for the virialized structures. Replacing the virialization by the turnaround condition would equip us with the statistical expectations for the masses of the structures at the turnaround as a function of redshifts within the same formalism. We leave the detailed examination of this concept to future work.  


The turnaround epoch, at most, belongs to the mildly non-linear regime, where the Lagrangian perturbation theory is expected to produce a reliable predictions. 
One advantage of the Lagrangian approach is that, also at the first order, it effectively reproduces higher orders in the standard Eulerian perturbation theory. 
This property validates the applicability of the analysis presented in this paper to the mildly non-linear scenarios~\cite{Peacock}. 
At the same time, the use of the RZA allows one to reproduce the results obtained within the Szekeres class II solutions as a particular case of our model, thus reinforcing its consistency. 
The fact that these formulas become exact for such general solution to Einstein's equations\footnote{
Szekeres models of class II are a generalization of the FLRW and Kantowski-Sachs solutions commonly used as a toy models for inhomogeneous universes~\cite{MIshak1,MIshak2,MeuresBruniPRD,MeuresBruniMNRas}.}
confirms that their predictions are a direct consequence of relativistic modelling.

It is worth noticing that the utility of the zero-average expansion regions extends beyond the applications proposed in this article. In particular, the zero-average expansion condition was proposed in \cite{Wiltshire2007} as an attempt to define the so-called finite infinity (\cite{Ellis1984}) i.e. a region that could effectively replace the spatial infinity in the context of isolated objects in the expanding Universe. The zero-expansion surface was examined in the context of an intimately related notion of the quasi-local energy in \cite{Faraoni2015}. For the non-virialized objects, however, the local zero-expansion surface lies well within the large-scale structure and thus is less relevant to the notion of finite infinity. 
For these reasons we claim that the zero-average expansion idea is potentially useful for cosmology and general relativity and deserves further examination.


\acknowledgments
The authors acknowledge the support of the National Science Centre (NCN, Poland) under the Sonata-15 research
grant UMO-2019/35/D/ST9/00342. We wish to thank Thomas Buchert and Nezihe Uzun for many valuable comments on this manuscript.


\bibliographystyle{plain} 
\bibliography{references} 
\end{document}